\documentclass[showpacs,prb,twocolumn]{revtex4}
\usepackage{epsf}
\usepackage{graphicx}
\usepackage{multirow}
\date{\today}

\newcommand{\wz}{\omega^0}
\newcommand{\bs}{{\mathbf S}}
\newcommand{\bq}{{\mathbf q}}
\newcommand{\bk}{{\mathbf k}}
\newcommand{\bP}{{\mathbf P}}
\newcommand{\ham}{{\mathcal H}}

\begin{document}
\title{Spinon-holon interactions in an
anisotropic $t$-$J$ chain: a comprehensive study} 
\author{Jurij \v{S}makov, A. L. Chernyshev, Steven R. White}
\affiliation{Department of Physics and Astronomy, University of
California Irvine, Irvine, California 92697, USA}
\begin{abstract}
We consider a generalization of the one-dimensional $t$-$J$ model with
anisotropic spin-spin interactions. We show that the anisotropy leads
to an effective attractive interaction between the spinon and holon
excitations, resulting in a localized bound state. Detailed
quantitative analytic predictions for the dependence of the binding
energy on the anisotropy are presented, and verified by precise
numerical simulations. The binding energy is found to interpolate
smoothly between a finite value in the $t$-$J_z$ limit and zero in
the isotropic limit, going to zero exponentially in the vicinity of
the latter. We identify changes in spinon dispersion as the primary
factor for this non-trivial behavior.
\end{abstract}
\maketitle
\section{Introduction}

One-dimensional (1D) lattice models of strongly correlated fermions
and bosons have traditionally been an object of intense theoretical
studies. The reason for such interest is twofold. First, such models
are relevant for the description of many real physical systems, such
as materials with strong uniaxial anisotropy, optical lattices, and
quantum nanowires. Second, there is a number of theoretical methods,
unique to one dimension, which allow either exact (solution via Bethe
Ansatz) or quasi-exact (bosonization, various renormalization group
schemes) treatment of the models in question.\cite{giamarchi-1d}

Out of the vast variety of 1D models of strongly correlated fermions,
the one known as the $t$-$J$ model clearly stands out as simple, yet
remarkably versatile. It captures both the ability of the particles to
hop from one site to another, and the spin-spin interactions between
them. By tuning the ratio of the coupling constants and the doping
level, it may be used to describe many 1D systems, ranging from
non-interacting mobile fermions to Heisenberg spin chains.
Furthermore, it also represents a physically relevant limit of another
1D model of paramount importance -- the Hubbard model.

In the one-dimensional $t$-$J$ model spin and charge dynamics are
independent, leading to the well-known effect of spin-charge
separation:\cite{LiebWu} the splitting of the electron (hole)
into spinon and holon elementary excitations that carry only spin and
only charge, respectively. This may be observed already at the
single-hole doping level. In that case the low-energy spectrum of the
$t$-$J$ model has been extensively studied in the
past.\cite{shiba-ogata-review,bares-tj-exact-2t,sorella-1d,muramatsu,Bernevig} Recently
it has been also shown that spinon and holon excitations are affected
by effective attractive interaction which, however, does not result in
their binding or pairing.\cite{Bernevig} This seeming controversy
encouraged us to explore in detail the nature of spinon-holon
interactions. In this paper we consider the $t$-$J$ model as a
limiting case of a more general model, which has anisotropic
($XXZ$-like) spin-spin interactions. In this model the effective
attractive spinon-holon interactions naturally emerge, leading to a
spinon-holon bound state. We present detailed quantitative analytical
predictions for the behavior of the binding energy as a function of
anisotropy and the implications of this physical picture for the
isotropic case, and verify them with precise numerical simulations,
using exact diagonalization (ED) and density-matrix renormalization
group (DMRG) on systems of up to $23$ and $128$ sites, respectively. 
An experimental test of our work could come
from photemission studies in insulating spin-chain materials with the Ising anisotropy, such as CsCoCl$_3$, CsCoBr$_3$, and others.\cite{nagler-1d-ising} 
We note that in the past such experiments in the {\it isotropic} Heisenberg 
spin-chain material of the cuprate family, SrCuO$_3$, have provided 
direct experimental evidence of spin-charge separation 
in the real $t$-$J$ model-like system.\cite{kim-separation-srcuo2} 

The one-dimensional $t$-$J$ model is defined by a Hamiltonian
${\mathcal H}=\sum_{i}{\mathcal H}_{i,i+1}$ with
\begin{equation}
\label{hamtj}
{\mathcal H}_{ij}=-t\sum_{\sigma}(c^{\dagger}_{\sigma
  i}c^{\phantom{\dagger}}_{\sigma j}+\text{H.c.}) 
+J\left(\bs_i\cdot\bs_{j}-\frac{n_in_j}{4}\right),
\end{equation}
where $c_{\sigma i}$ annihilates a fermion with spin $\sigma$ on site
$i$, $n_i$ is the fermion number operator on site $i$, and $\bs_i$ is
the fermion spin operator. Periodic boundary conditions (BCs) are
assumed. The Hilbert space, in which the Hamiltonian (\ref{hamtj})
acts, is restricted to a subspace without any doubly-occupied
sites. At half-filling (one fermion per site) no particle hopping is
possible, so the model is reduced to an isotropic Heisenberg model of
interacting spins, with an antiferromagnetic (AF) ground state (GS).
Doping it, even with a single hole, leads to spin-charge separation,
which is manifested by the splitting of quasiparticle peaks in the
excitation spectrum into two different sets, with energies scaling
with $t$ or $J$, respectively.\cite{LiebWu}

In order to study the spinon-holon interaction as a function of
anisotropy, we consider a generalization of the $t$-$J$ model
(\ref{hamtj}) with a Hamiltonian ${\mathcal H}_{ij}$ of the form
\begin{eqnarray}
\label{hamtjp}
{\mathcal H}_{ij}&=&-t\sum_{\sigma}(c^{\dagger}_{\sigma
  i}c^{\phantom{\dagger}}_{\sigma j}+\text{H.c.})\nonumber\\ 
&+&J_z\left(S^z_i
  S^z_j+\alpha\bs^{\perp}_i\cdot\bs^{\perp}_{j}-\frac{n_in_j}{4}\right). 
\end{eqnarray}
Here $\bs^{\perp}_i\cdot\bs^{\perp}_j=S^x_iS^x_j+S^y_iS^y_j$, and
parameter $\alpha$ controls the anisotropy of spin-spin interactions.
The original isotropic $t$-$J$ model is recovered by setting $\alpha=1$.

\begin{figure}[b]
\includegraphics*[width=0.8\hsize,scale=1.0]{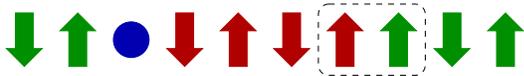}
\caption{(Color online). 
A hole  in the Ising AF background (circle), moved by four sites from origin.
The location of immobile spinon is indicated by the dashed box. 
\label{fig:spinon_holon}
}
\end{figure}

The $\alpha=0$ limit of Hamiltonian (\ref{hamtjp}) is known as
$t$-$J_z$ model. Its GS in the undoped state is an Ising
antiferromagnet, and the effect of doping it with a single hole is
easy to understand (see Fig. \ref{fig:spinon_holon} for an
illustration). It results in a creation of a spinon-holon bound state
due to an effective attraction between the immobile spinon (the
Hamiltonian does not contain any spin-flipping term which would allow
it to propagate) and a free holon.\cite{short} The binding
energy can then be calculated analytically:
\begin{equation}
\label{ising-delta}
\Delta = 2t\left[1-\sqrt{1+(J_z/4t)^2}\right].
\end{equation}
We present several different methods to obtain this result in the Appendix.
Setting $\alpha$ to a non-zero value presents three distinct
possibilities.  First of all, it is possible that any non-zero value
of $\alpha$ immediately destroys the bound state, so $\alpha=0$ is the
only singular point in the phase diagram with a finite $\Delta$.
Second, there is a possibility that $\Delta$ varies smoothly with
$\alpha$, interpolating between the finite value at $\alpha=0$ and
zero value in the isotropic case. Finally, the $\Delta(\alpha)$
dependence can go to zero at some non-trivial critical value $0<
\alpha_c < 1$. Out of these possibilities the first one appears to be
the least likely one, as it is intuitively clear that small
transverse spin-spin interaction cannot immediately destroy the bound
state. While at $\alpha \not = 0$ the spinon will become mobile, for
small $\alpha$ it is still going to be too ``massive'', compared to
virtually free holon. We cannot unequivocally rule out the last option
(binding becomes too weak to be detected numerically near the
isotropic limit), but we argue that in this regime the spin background
is Ising-like, with long-range spin order for any anisotropy $\alpha <
1$. This fact strongly suggests that the only anisotropy-driven
critical point in the system is at $\alpha=1$. To confirm this
hypothesis and carefully examine the remaining option, a detailed
investigation of the binding energy as a function of $\alpha$ is
required. Such an investigation is the main topic of this paper.

We have chosen the representative value of $J_z/t=4.0$ for most of our
calculations, after confirming that the results at other values of
$1\le J_z/t\le 8$ are qualitatively similar. We also present the final
results for the binding energy for $J_z/t=1.0$. In general, we
do not expect any qualitative difference for any other $J_z/t$ value
as Eq. (\ref{ising-delta}) gives $\Delta<0$ for any $J_z$.
The choice of $J_z/t=4.0$ was made mostly to optimize the numerical 
accessibility of the binding energy in a wider range of $\alpha$.

The rest of the paper is organized as follows. In section
\ref{sec:numerics} we present our numerical results, discussing in
detail the finite-size effects of the data and the procedure for
extrapolation to the infinite system size. Section \ref{sec:theory}
contains the theory for the binding energy, based on Bethe-Salpeter
equation. We summarize our results in section \ref{sec:conclusions},
and present three different ways to derive the expression for the
binding energy of $t$-$J_z$ model (\ref{ising-delta}) in the Appendix.

\begin{figure}[t]
\includegraphics*[width=1.0\hsize,scale=1.0]{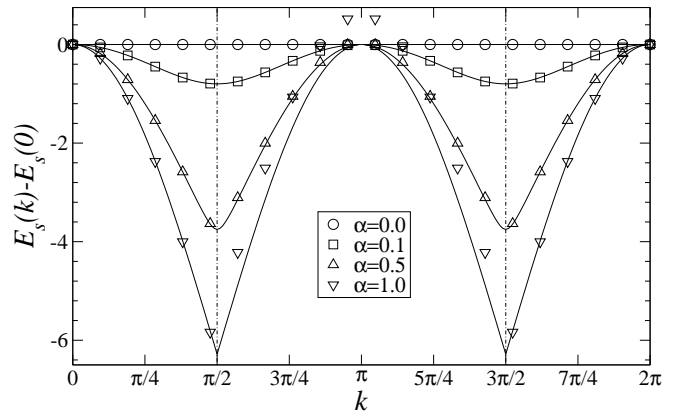}
\caption{Lowest ED energies $E_s(k)-E_s(0)$ vs $k$ in $L=21$ chain
 with zero holes at $J_z/t=4.0$ (spinon dispersion).  Solid lines show
 spinon dispersion for an infinite system obtained from
 BA.\cite{johnson-excitations-xyz} All the energies are in units of $t$.
\label{fig:spinon_dispersion}
}
\end{figure}
\begin{figure}[b]
\includegraphics*[width=1.0\hsize,scale=1.0]{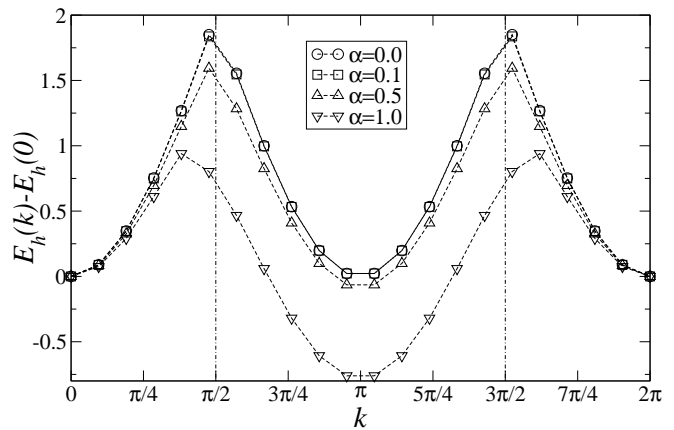}
\caption{Lowest ED energies $E_h(k)-E_h(0)$ 
 vs $k$ in $L=21$ chain with one hole at $J_z/t=4.0$ (holon dispersion). 
Dashed lines are guides to the eye. All the energies are in units of $t$.
\label{fig:holon_dispersion}
}
\end{figure}

\section{Numerical results}
\label{sec:numerics}
We have used the ED and DMRG techniques to calculate the ground state
energies (GSEs) of the model for different system sizes and doping
levels. This information was then used to extract the binding energy
of a spinon-holon state in the infinite size limit.

In ED we start by considering a subset of states of a system of size
$L$ with given total hole number $n$ and total $S^z$. To take
advantage of the translational symmetry, we then use these to
construct a basis out of eigenstates of the translation operator with
a given momentum $k$. Finally, the Hamiltonian matrix in this reduced
basis is constructed, and its lowest eigenvalue is calculated
iteratively using the Lanczos algorithm. The implementation of every step
in the procedure is described in detail in
Ref. \onlinecite{haas}. With these techniques we were able to
calculate the GSEs of systems of up to 23 sites with ED. Using
DMRG\cite{white-one,white-two} we have calculated GSEs of systems of
up to $L=128$ sites using periodic boundary conditions (PBCs), which
greatly increases the numerical effort required. Up to $m=1400$ states
per block were kept in the finite system method, with corrections
applied to the density matrix to accelerate convergence with
PBCs.\cite{white-three}

We have carefully tested our algorithms by comparing the results of ED
and DMRG for different system sizes, both with and without the
hole. We have also compared the ED energies with the independent
results for the GSEs of the $XXZ$ model. \cite{medeiros-size-corr} In
all cases agreement to at least 7 decimal places was achieved.

The elementary excitations of the model may be studied by looking at
systems of different sizes with either no or one hole. In the case of 
an odd
number of sites and no holes the PBCs are frustrating, corresponding
to creation of a frustrated ferromagnetic link -- a spinon
excitation. The lowest energies for each momentum sector for a chain
of 21 sites with PBC at different anisotropies are shown in
Fig. \ref{fig:spinon_dispersion}. This gives us the spinon
dispersion, which evolves from completely flat in the Ising case
$\alpha=0$ to quasi-relativistic in the isotropic case
$\alpha=1$. Solid lines in Fig. \ref{fig:spinon_dispersion} show the
exact BA result for the spinon spectrum in the
$XXZ$-model:\cite{johnson-excitations-xyz}
\begin{equation}
\label{spinon_dispersion}
\omega_q\!=\!c\sqrt{1-\kappa^2\sin^2q}.
\end{equation}
Here $c/J_z=K\sqrt{1-\alpha^2}/\pi$, and $\kappa$ is determined
from the condition $\pi K'/K=\cosh^{-1}\left(1/\alpha\right)$,
where $K\equiv K(\kappa)$ and $K'\equiv K(\sqrt{1-\kappa^2})$
are complete elliptic integrals of the first kind. The lowest
spinon energy is attained at $q=\pm\pi/2$ for any $\alpha > 0$.
\begin{figure}[t]
\includegraphics*[width=1.0\hsize,scale=1.0]{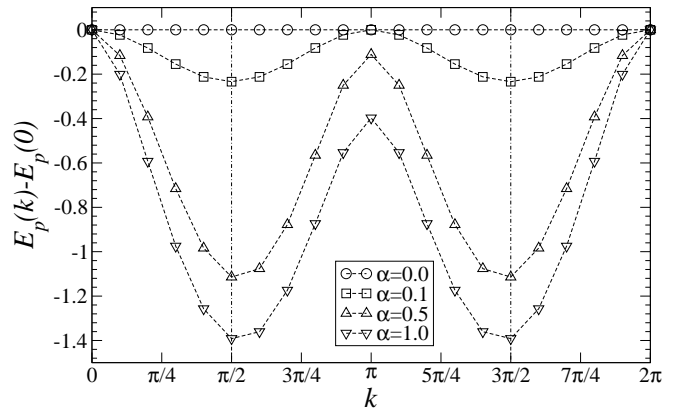}
\caption{Lowest ED energies $E_p(k)-E_p(0)$ vs $k$ in $L=20$ chain
 with one hole at $J_z/t=4.0$ (spinon-holon pair dispersion).  Dashed
 lines are guides to the eye.
\label{fig:pair_dispersion}
}
\end{figure}

A configuration with odd number of sites and one hole contains a
``pure'' holon, which can propagate through the system without
disturbing the otherwise perfect AF background. Typical holon
dispersions, obtained by measuring the energies of a 21-site chain
with one hole, are presented in Fig. \ref{fig:holon_dispersion}.  The
holon's minimum energy dependence on $\alpha$ is non-trivial.  At
$\alpha=0$ the system has unique lowest energy point at momentum
zero. This is only true for a finite system though, as in the infinite
system this energy would be degenerate with the one at
$k=\pi$. However, since we do not have a reciprocal space point
exactly at $k=\pi$, for a finite system the energy at the momentum
points closest to $k=\pi$ is somewhat higher. This mismatch is an
important source of finite-size effects in our measurement, as we
discuss below. As the anisotropy $\alpha$ is increased, the energy at
$k=\pi$ decreases, so the ground state switches from the $k=0$ to
$k=\pi$ sector at some finite intermediate value. Notably, these
observations are in stark contrast with assumptions by Shiba and
Ogata,\cite{shiba-ogata-review} who claim that the $k=0$ and $k=\pi$
energies are going to be degenerate for any finite system in the
isotropic case (they are degenerate in an infinite system
though). Furthermore, their interpretation of the holon dispersion
(presented in Fig. 6 of Ref. \onlinecite{shiba-ogata-review}) is
somewhat misleading: they attribute the double-peaked structure of the
dispersion to ``strong antiferromagnetic correlations''. Our
calculations confirm that the holon dispersion close to the $k=0$ and
$k=\pi$ points may be very well fitted with a simple cosine dispersion
of a free particle. This indicates that the characteristic
double-peaked shape is formed by two different holon branches,
centered at $k=0$ and $k=\pi$. As one moves away from these points
towards $k=\pi/2$, the energy of the excitations grows, eventually
making the creation of a spinon-antispinon pair energetically
favorable, as suggested in Ref. \onlinecite{Bernevig}. That results in
mixing of the two holon branches, which leads to a rounding of the
dispersion peaks.
\begin{figure}[b]
\includegraphics*[width=\hsize,scale=1.0]{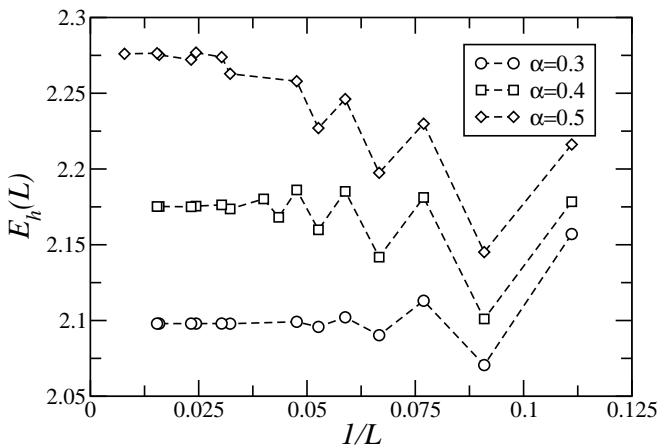}
\caption{Raw holon GS energies for $J_z/t=4.0$ and different anisotropies
$\alpha$ as a function of inverse system size $1/L$. Dashed lines
are guides to the eye.
\label{fig:size_mess}
}
\end{figure}

Finally, an even-sized system with one hole corresponds to a situation
where both spinon and holon are present. The GSE as a function of $k$
for the system containing a spinon-holon pair is presented in
Fig. \ref{fig:pair_dispersion}.

We can measure the energies of an interacting spinon-holon pair, as
well as those of individual spinon and holon excitations, by taking
the GSE of a corresponding configuration and subtracting the extensive
part of the energy $\tilde{\epsilon}_\alpha L$, where
$\tilde{\epsilon}_\alpha$ is the energy per site of an infinite $XXZ$
chain, known from Bethe Ansatz.\cite{yang-xxz} That way we can obtain
the spinon, holon, and spinon-holon pair energies for a set of
different system sizes. After extrapolating to the infinite system
size, the corresponding energies $E_s$, $E_h$, and $E_p$ can be used to
calculate the binding energy of the spinon-holon state in an infinite
system as
\begin{equation}
\label{bindinga}
\Delta=E_p-E_s-E_h.
\end{equation}
We will refer to this approach as ``method $A$''.
\begin{figure}[t]
\includegraphics*[width=1.0\hsize,scale=1.0]{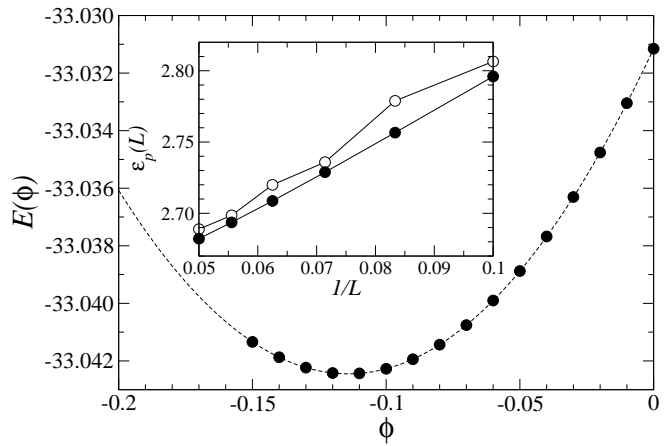}
\caption{Ground state energy as the function of phase $\phi$ for
$L=16$, $J_z/t=4.0$, $\alpha=0.5$. Dashed line is a quadratic fit. Inset
shows the comparison of raw staggered pair energy data (open circles)
and phase-corrected data (solid circles) for the same $J_z$ and
$\alpha$, and different system sizes.
\label{fig:phase_dep}
}
\end{figure}
\begin{table}[b]
\begin{tabular}{|c|c|c|}
\hline
Branch & $\text{mod}(L-1,4)$ & $\text{sign}(t)$\\
\hline
\multirow{2}{*}{$4$-even} & 0 & +1 \\
& 2 & -1 \\
\hline
\multirow{2}{*}{$4$-odd} & 0 & -1 \\
& 2 & +1 \\
\hline
\end{tabular}
\caption{
Splitting of holon energy data for different sizes $L$ 
and different signs of the hopping constant $t$ into the
$4$-even and $4$-odd branches.
\label{tab:branches_simple}
}
\end{table}

In order to obtain an accurate estimate for the excitation energies in
the infinite size limit, we have to deal with a variety of finite-size
effects. The lifting of degeneracy in holon dispersion mentioned above
is one of them. It turns out that its effect on the resulting GSE
depends on whether the system size $L$ (or $L-1$ if $L$ is odd) is
divisible by 4 or not, so we will refer to these two data branches as
$4$-even and $4$-odd, respectively. Such a $\textrm{mod}(4)$
dependence has been extensively discussed in the literature (see
Ref. \onlinecite{Bernevig} and references therein).  In the $4$-even
branch the energy $\epsilon_{k=0}$ provides an upper bound for the
true GSE, while $\epsilon_{k=\pi}$ serves as a lower bound, and the
bounds are reversed for the $4$-odd case.  Another source of finite-size
corrections is the incommensurability of the momentum space points in
the systems of odd size. For example, it can be seen from
Fig. \ref{fig:spinon_dispersion} that for spinons the GS corresponds
to momentum $\pi/2$ in the $L=\infty$ limit. However, for any
finite-sized system with odd $L$ there will be \emph{no} reciprocal
space point $k=\pi/2$, instead the GSE will occur at one of the
nearest points with momentum $k=\pi/2\pm \delta_L$, where with
$\delta_L=\pi/L$. As the system size is increased, $\delta_L$ will go to
zero, and the GSE will drift towards its infinite-$L$ limiting value.
Similarly, we cannot directly measure the GSE for holons
(Fig. \ref{fig:holon_dispersion}) at $k=\pi$. All these factors lead
to a highly non-trivial finite size dependence. As an example,
Fig. \ref{fig:size_mess} shows the size dependence of the raw holon
energies. To get a meaningful extrapolation the separate analysis of
$4$-even and $4$-odd branches, which contain only half of the original
points, is required.
\begin{figure}[b]
\includegraphics*[width=\hsize,scale=1.0]{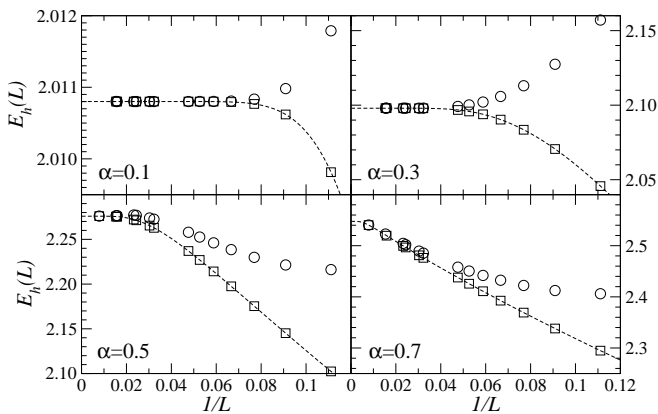}
\caption{The $4$-even (circles) and $4$-odd (squares) branches of
holon energy data for $J_z/t=4.0$ and different anisotropies $\alpha$.
The ``good'' branch is used to extrapolate to $L=\infty$, using form
(\ref{fitform}) with $n=4$ and the WdV values of $\lambda$ from
Ref. \onlinecite{vega-woynar-size-corr}
\label{fig:holons}
}
\end{figure}

The situation with incommensurate $k$-space points can be improved by
imposing twisted boundary conditions on the model.\cite{zotos-tbc} A
boundary twist leads to the translation of the points in $k$-space,
but does not affect the energy spectrum. Thus, by adjusting the twist
one can shift the $k$-point with anticipated minimum energy from an 
incommensurate location in $k$-space to an accessible one. In case of
holons such a procedure is particularly simple, since we have to use a
phase shift of $\pi$ to move the $k=\pi$ point of the original model
to momentum $k=0$ for a model with boundary twist. Such a phase shift
is readily implemented just by switching the sign of the hopping
constant $t$ to the opposite one. That enabled us to reconstruct the
points, lost due to the splitting into $4$-even and $4$-odd branches,
by complementing the holon GSE data with measurements performed on the
model with $t=-1$, as shown in Table \ref{tab:branches_simple}.

The spinon-holon pair GSE data also suffer from the $k$-mismatch, as
the GS is achieved at an incommensurate $k$-point.\cite{zotos-tbc} In
principle, the same procedure may be applied to improve the pair
energy data. There, however, the phase shift $\phi$ needed to shift
the energy minimum to an accessible momentum point is size-dependent,
so it has to be determined individually for every data point. By
replacing $t$ in (\ref{hamtjp}) by $te^{i\phi}$ and tuning the phase
shift $\phi$, we were able to measure the total energy of the system
as a function of $\phi$ using ED. An example of such dependence is
presented in Fig. \ref{fig:phase_dep}. One remarkable feature of this
dependence is that it is very well fit by a quadratic polynomial, so
it is sufficient to know the energy at two different non-zero values
of $\phi$ to recover the ``true'' lowest energy at the minimum with
excellent accuracy. The inset of Fig. \ref{fig:phase_dep} shows
dramatic improvement of the data due to the phase-induced correction.
While such boundary conditions can be readily handled by ED, our DMRG
code required extensive modifications to support them. 
Thus, in this work we perform the extrapolations
using only the raw spinon-holon pair GSE data, split into $4$-even and
$4$-odd branches. Further improvement 
of the precision of our results by using phase-adjusted
data is possible. 
\begin{figure}[t]
\includegraphics*[width=\hsize,scale=1.0]{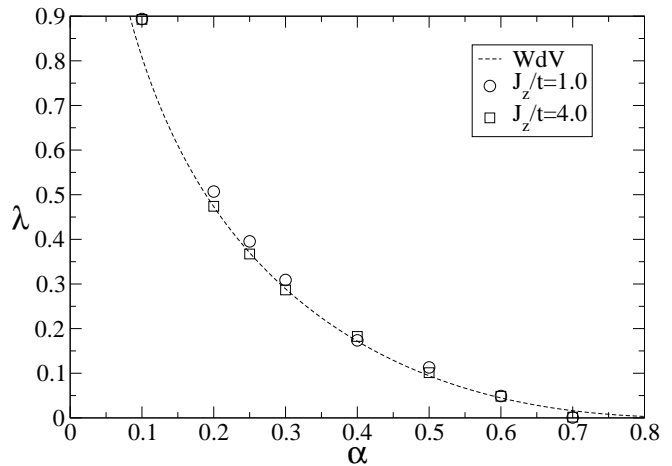}
\caption{Comparison of the analytic values of $\lambda$ from
Ref. \onlinecite{vega-woynar-size-corr} (dashed line) with the values
obtained by fitting the holon data with form (\ref{fitform}) and
keeping $\lambda$ a free fitting parameter (symbols).
\label{fig:lambdas}
}
\end{figure}

\begin{figure}[b]
\includegraphics*[width=\hsize,scale=1.0]{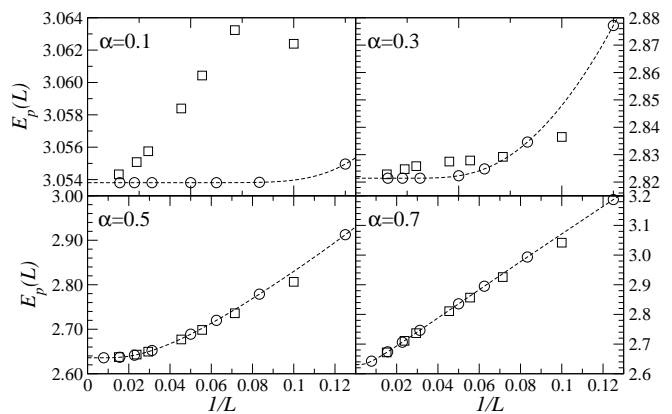}
\caption{The $4$-even (circles) and $4$-odd (squares) branches of
spinon-holon pair energy data for $J_z/t=4.0$ and different
anisotropies $\alpha$.  The ``good'' branch is used to extrapolate to
$L=\infty$, using form (\ref{fitform}) with $n=4$.
\label{fig:pairs}
}
\end{figure}

After the data for holon and pair are split into such branches, we
need to extrapolate them to the $L=\infty$ limit, using a reasonable
fitting form. From the holon (Fig. \ref{fig:holons}) and pair
(Fig. \ref{fig:pairs}) excitation energy data it is evident, that it
has a complicated size dependence, which cannot be adequately
described by a polynomial. Clearly, at large $L$ the difference
between the limiting value and the data points drops exponentially
with increasing $L$. Incidentally, the size dependence for the GSE of
the $XXZ$ model, deduced by Woynarovich and de Vega (WdV) from
BA,\cite{vega-woynar-size-corr} is also dominated by an exponential
factor $\exp(-\lambda L)$. That inspired us to attempt fitting the
holon and pair data with the functional form
\begin{equation}
\label{fitform}
E_L=E_{\infty}+e^{-\lambda L}P_n(1/L),
\end{equation}
where $P_n(x)$ is a polynomial of order $n$ ($n\le 4$) in $x$ with
adjustable coefficients. Not only does it work remarkably well
for both holons and pairs (extrapolations are shown in
Fig. \ref{fig:holons} and Fig. \ref{fig:pairs} with dashed lines), but
the values of the coefficient $\lambda$ we have found by keeping this
parameter free in our holon fits provide an excellent match to the
analytic values found by WdV for the $XXZ$ model (their comparison is
presented in Fig. \ref{fig:lambdas}). Therefore, we have assumed that
for holons the $\lambda$ values found by WdV are either exact, or a
very good approximation. Thus, we used them in our holon fits,
reducing the total number of free parameters by one.  For pairs the
values of the $\lambda$ parameter did not correlate with WdV
results at all, so it had to be kept as a free parameter in the fit.

When the parameter $\lambda$ is sufficiently large (for $\alpha \lesssim
0.3$), the exponential factor in (\ref{fitform}) makes the asymptotic
approach to the infinite value very rapid, allowing us to simply adopt
the energy value for the largest available size as the infinite-size
limiting value. Increasing $\alpha$ results in decreasing $\lambda$,
which pushes the onset of the exponential size dependence to larger
and larger system sizes. In this regime the extrapolation using form
(\ref{fitform}) must be used. Around $\alpha \sim 0.5$ parameter
$\lambda$ becomes comparable with the inverse of the maximum available
system size. At higher anisotropies the onset of the exponential
behavior in size dependence takes place at characteristic sizes, not
accessible by our calculations (as can be seen on the lower right
panel of Fig. \ref{fig:pairs}), making the precise extrapolation of
the pair excitation energy impossible. We can improve the accuracy of
the extracted infinite size value $E_\infty$ by noting that both for
holons and pairs one of the branches is always more ``well-behaved''
than the other one. For example, non-uniform behavior of the $4$-even
branch for holons can be seen on the lower left panel of
Fig. \ref{fig:holons} (it peaks slightly around $1/L=0.025$), and on
the upper panels of Fig. \ref{fig:pairs} for the $4$-odd pair
branch. This non-uniformity of the ``bad'' branch usually results from
the GS switching from one momentum sector to a different one as a
function of $L$. In our analysis we have used only the extrapolations
obtained with the ``well-behaved'' branch -- $4$-odd for holons and
$4$-even for pairs. The energy for the spinon excitations can, in
principle, be extracted from the numerical data in a similar
way. However, to further improve our results, we have used the
analytic expression for the spinon excitation energy $E_s=\omega_{q=\pi/2}$, given
by Eq. (\ref{spinon_dispersion}), thus eliminating the finite-size effects
from the spinon GSE completely.
\begin{table}[t]
\begin{tabular}{|c|c|c|c|}
\hline
Branch & $\text{mod}(L,4)$ & $\text{sign}(t)$ for $E^1_{L-1}$ & $\text{sign}(t)$ for $E^1_{L+1}$\\
\hline
$B1$ & 0 & -1 & +1 \\
$B2$ & 0 & +1 & -1 \\
$B3$ & 2 & -1 & +1 \\
$B4$ & 2 & +1 & -1 \\
\hline
\end{tabular}
\caption{
Subdivision of the numerical data for $\Delta(L)$ into different branches
due to finite size effects in method $B$.
\label{tab:branches_b}
}
\end{table}
\begin{figure}[b]
\includegraphics*[width=1.0\hsize,scale=1.0]{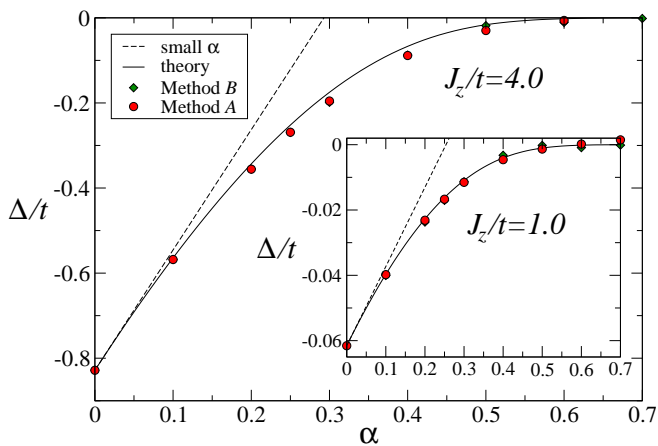}
\caption{(Color online). Binding energy $\Delta$ as a function of $\alpha$ for
$J_z/t=4.0$ and $J_z/t=1.0$ (inset). Data includes the theoretical
prediction (solid line), numerical results from ED and DMRG data
obtained by method $A$ (circles) and method $B$ (diamonds). Dashed
line shows the linear approximation (\ref{linear_small}), valid at small
$\alpha$.
\label{fig:binding_both}
}
\end{figure}

Finally, the binding energy results for $L=\infty$ 
obtained by method $A$ using (\ref{bindinga}) for $J_z/t=4.0$ and
$J_z/t=1.0$ are presented in Fig.  \ref{fig:binding_both}. 
These data are of high-precision for $\alpha < 0.5$.
We estimate the maximum relative error of the resulting
binding energy by studying the quality of the
fits and the variation of $\Delta_\infty$ depending on the fit
type. At $\alpha=0.5$ the error does not exceed 3\% (10\%) for $J_z/t=4.0$ ($J_z/t=1.0$) 
and becomes negligible very rapidly for smaller values of $\alpha$. 
For $\alpha=0.6$ the error is of the order of 10\% (100\%)
for $J_z/t=4.0$ ($J_z/t=1.0$), and for 
$\alpha=0.7$ it exceeds $100$\% for both representative values of $J_z$.
As mentioned before, we do not
consider the binding energy results for $\alpha > 0.6$ to be reliable
due to issues with pair energy extrapolation. Also, at larger values
of $\alpha$ the value of the binding energy becomes comparable with
the accuracy of our DMRG method (about $10^{-7}$ to $10^{-8}$ absolute
precision, leading to about $10^{-4}t$ accuracy of the binding energy
at $L=128$ and $J_z/t=4.0$), imposing a natural limitation on the
quality of the data.
\begin{figure}[t]
\includegraphics*[width=1.0\hsize,scale=1.0]{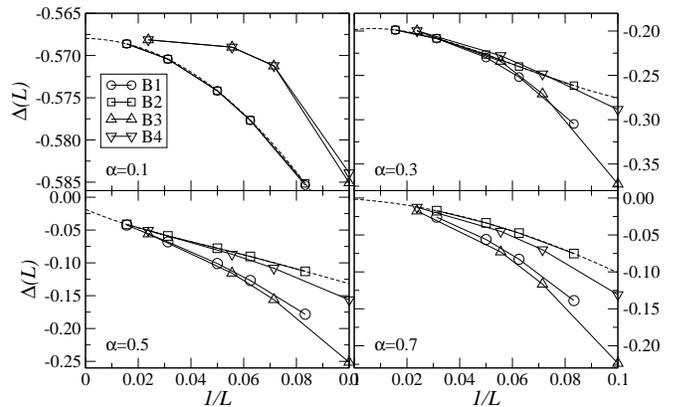}
\caption{Size dependence for different branches of binding energy
$\Delta(L)$ at $J_z/t=4.0$. For branch definition see Table
\ref{tab:branches_b}. Solid lines are guides to the eye, dashed curve
is the polynomial extrapolation of the ``good'' branch $B2$. Legend
applies to all panels.
\label{fig:branches}
}
\end{figure}

An alternative way to extrapolate the binding energy to the infinite
size limit is to calculate it for every system size $L$ individually,
and then do the extrapolation of the resulting size dependence to
$L=\infty$. In this method (referred to as ``method $B$'') we
intentionally avoid using any BA results, to see whether the reliable
binding energy data may be obtained based on the numerical results
alone. One could hope that the finite-size effects of various
components entering the binding energy may cancel out, allowing the
extrapolation to the infinite-size limit using a simple polynomial in
$1/L$, instead of an exponential. This approach, not depending on the
theoretical results, provides an important validity test for the
results of method $A$.

Since holon and spinon GSEs are only available for odd $L$, and the
spinon-holon pair ones only for even $L$, we define the finite size
binding energy for an \emph{even} size $L$ as
\begin{equation}
\Delta(L)=E^0_L+E^1_L-[E^0_{L-1}+E^0_{L+1}+E^1_{L-1}+E^1_{L+1}]/2,
\end{equation}
where $E^h_L$ is the ground state of a system with $L$ sites, doped
with $h$ holes. This expression is analogous to (\ref{bindinga}): sum
of first two terms corresponds to the pair energy, while
$(E^0_{L-1}+E^0_{L+1})/2$ and $(E^1_{L-1}+E^1_{L+1})/2$ represent the
average energy of a system of size $L$ with a spinon and holon,
respectively. Again, due to staggering of the GSEs, binding energy
data splits into a $4$-even and $4$-odd branches, depending on whether
$L$ is divisible by $4$ or not. However, in this case we have an
additional freedom of choosing the sign of $t$ for holon energies
$E^1_{L-1}$ and $E^1_{L+1}$. Taking this into account results in 4
different data branches, defined in Table \ref{tab:branches_b}.  The
remaining possibilities of using the same sign of $t$ both for
$E^1_{L-1}$ and $E^1_{L+1}$ have been discarded as obviously
suboptimal.

The size dependence of different data branches for different
anisotropies is presented in Fig. \ref{fig:branches}. Due to
splitting, the number of points in each branch is pretty small, so we
have used a polynomial of maximum possible degree (one less than the
number of points) to perform the extrapolation to the infinite system
size. From our previous experience we know that the ``good'' holon
branch in method $A$ corresponds to branch $B2$, therefore we used the
extrapolated value from this branch as our final result for the
binding energy. As can be seen in Fig. \ref{fig:binding_both}, the
data from methods $A$ and $B$ are in excellent agreement in the range
of $\alpha$, where its calculation is reliable.  However, we have
found that the method $B$ data always has a larger relative error than
method $A$, mainly due to the size-dependence of the spinon component,
eliminated in method $A$.

\section{Theoretical results}
\label{sec:theory}
Because in the Ising limit the spinon is impurity-like,
the spinon-holon binding at $\alpha=0$ can be solved in a number of
ways (see Appendix) to give Eq. (\ref{ising-delta}).  For finite
$\alpha$ the binding energy of the spinon-holon state may be
calculated analytically by finding the poles of the two-particle
scattering amplitude $\Gamma$. We will use shorthand notation $\bq=(q,
\omega)$ and $\bk=(k, \epsilon)$ to denote the momenta and energies of
spinon and holon, respectively.
\begin{figure}[b]
\includegraphics*[width=1.0\hsize,scale=1.0]{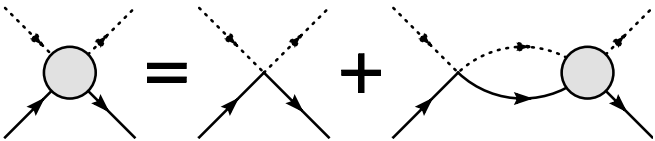}
\caption{Bethe-Salpeter equation for the spinon-holon scattering
amplitude (circle). Spinons (holons) are
shown by dashed (solid) lines.
\label{fig:diagram}
}
\end{figure}

Scattering amplitude $\Gamma$ obeys the Bethe-Salpeter
equation,\cite{landau-qed} presented in diagrammatic form in
Fig. \ref{fig:diagram}. Generally, it depends on both the incoming
$\bq$, $\bk$ and outgoing $\bar{\bq}$, $\bar{\bk}$ $2$-momenta of
spinon and holon. Using that momentum and energy are
conserved, $\bk+\bq=\bar{\bk}+\bar{\bq}=\bP\equiv(P,E)$, we may write
it as
\begin{eqnarray}
\Gamma_{\bP}(\bq,\bar{\bq})=V_{q,\bar{q}}+
\int_{\bq'}
V_{q,q'}G^s_{\bq'}
G^h_{\bP-\bq'}\Gamma_{\bP}(\bq',\bar{\bq}).
\label{maing}
\end{eqnarray}
Here $V_{q,q'}$ is the spinon-holon interaction, and
$G^{h(s)}$ is the holon (spinon) Green's function.
A shorthand notation $\int_{\bq'}\equiv
\int_{-\infty}^\infty\,\frac{d\omega'}{2\pi}\,\sum_{q'}$, with 
$\sum_{q'}=\int_{-\pi}^\pi\frac{dq'}{2\pi}$ is used.
In the vicinity of the pole of $\Gamma$, Eq. (\ref{maing}) should reduce
to a homogeneous integral equation with  
$\Gamma$ whose dependence on one of the momenta
$\bar{\bq}$ is only parametric and can be
dropped\cite{landau-qed}
\begin{equation}
\label{homo}
\Gamma_{\bP}(\bq)=\int_{\bq'}\,V_{q,q'}G^s_{\bq'}
G^h_{\bP-\bq'}\Gamma_{\bP}(\bq').
\label{main_homo}
\end{equation}
The holon and spinon create a bound state if this integral equation
has a solution. 
 By introducing a function
\begin{equation}
\chi_{\bP}(q)=\int_\omega G^s_{\bq}G^h_{\bP-\bq}\Gamma_{\bP}(\bq),
\end{equation}
multiplying both sides of (\ref{homo}) by $G^s_{\bq}G^h_{\bP-\bq}$,
and integrating over $\omega$, we arrive at
\begin{equation}
\chi_{\bP}(q)=\left[\int_\omega G^s_{\bq}G^h_{\bP-\bq}
\right]
\sum_{q'} V_{q,q'}\chi_{\bP}(q')
.
\end{equation}
Evaluation of the first integral on the rhs requires knowledge of the
spinon and holon Green's functions. For now we will just assume that
they are free particles with some dispersions $\omega_q$ and
$\epsilon_k$, the specific form of which is to be determined:
\begin{eqnarray}
\label{gfs}
G^s_{q,\omega}&=&\frac{1}{\omega - \omega_q+i\delta},\nonumber\\
G^h_{k, \epsilon}&=&\frac{1}{\epsilon - \epsilon_k+i\delta}\nonumber.
\end{eqnarray}
With this assumption the integral is trivially done, yielding the final form
of the Bethe-Salpeter equation for $\chi_{\bP}(q)$:
\begin{equation}
\label{schrod_pair}
\chi_{\bP}(q)=\frac{1}{E-\epsilon_{P-q}-\omega_{q}}
\sum_{q'} V_{q,q'}\chi_{\bP}(q').
\end{equation}
From this equation it is clear that $\chi_{\bP}(q)$ is nothing but the
pair wavefunction and the equation (\ref{schrod_pair}) is the
Schr\"odinger equation for it in integral form.  The pair energy $E$ may
be thought of as the binding energy $\Delta$, measured relative to the
lowest energies of the particles $\epsilon_0=min[\epsilon_k]$ and
$\omega_0=min[\omega_q]$:
\begin{equation}
E = \Delta+\epsilon_0+\omega_0.
\end{equation}
In the Ising limit 
$\epsilon_k=-2t\cos k$, 
$\epsilon_0=-2t$, $\omega_q\!=\!\omega_0\!=\!J_z/2$, and 
$V_{q,q'}=-\omega_0$, so Eq. (\ref{schrod_pair}) is readily solved
by
\begin{equation}
\chi_{\bP}(q)=\frac{C}{E-\epsilon_{P-q}-\omega_{0}},
\end{equation}
yielding a dispersionless ($P$-independent) bound state with
$\Delta$ given by (\ref{ising-delta}). From general considerations, the
binding energy in 1D should scale as $-V^2m$, where $V$ is interaction
strength and $m$ is the particle mass. In the Ising case this gives
$\Delta\!\sim\!-J_z^2/t$, in agreement with the exact result
(\ref{ising-delta}).

Away from the Ising limit (at nonzero $\alpha$) the physical picture
changes qualitatively. First of all, due to the 
spin-flips the spinon is no longer
stationary; it may propagate through the lattice and has a $\pm\pi/2$
momentum in the ground state. Second, the spinon-holon interaction
$V_{q,q'}$ changes. Finally, the holon dispersion is altered as well
and may acquire some ``dressing''. The changes in the latter, however,
should not affect the pairing in any significant way due to the
fact that only the holon dispersion near the energy minimum matters
for it. The holon mass renormalization has been analyzed in detail in
Ref. \onlinecite{zotos-tbc} and it was found insignificant throughout the
anisotropic regime $0\le \alpha \le 1$. This means that both the
``dressing'' and the holon dispersion changes are minor and should not
affect pairing. On the other hand, changes in the spinon dispersion
are qualitative and drastic. At $\alpha=0$, the spinon may be viewed as a
gapped, immobile excitation with the energy $\omega_q\!=J_z/2$.  With
increasing $\alpha$ it evolves into a relativistic one, turning
completely gapless in the isotropic limit $\alpha=1$, where its
dispersion is $\omega_q=J_z(\pi/2)|\cos q|$. The spinon dispersion for the
$XXZ$ model at intermediate values of $\alpha$, shown by solid lines
in Fig. \ref{fig:spinon_dispersion}, is known exactly from
BA,\cite{johnson-excitations-xyz} see
Eq. (\ref{spinon_dispersion}). While the parameter $c/J_z$ in this
equation changes almost linearly between $1/2$ and $\pi/2$ as $\alpha$
goes from $0$ to $1$, the parameter $\kappa$ varies from $0$ to $1$ rather
steeply, achieving the value of approximately $0.996$ at
$\alpha=0.5$. As a result, the spinon gap $\omega_s$ given by
$\omega_s=c\sqrt{1-\kappa^2}$ becomes sufficiently small already at
$\alpha\sim 0.5$. One can obtain the asymptotic behavior for
$\omega_s$, valid for $\alpha \gtrsim 0.5$, and show that it approaches zero
exponentially in $(1-\alpha)^{-1/2}$ as $\alpha \rightarrow 1$:
\begin{equation}
\label{gap_asymp}
\omega_s \approx
4c\exp\left(-\pi^2\sqrt{\frac{\alpha}{8(1-\alpha)}}\right).
\end{equation}
The smallness of the spinon gap may be used to write the spinon spectrum
in approximate ``quasi-relativistic'' form in this regime:
\begin{equation}
\omega_q = \sqrt{c^2\left(q-\frac{\pi}{2}\right)^2+\omega_s^2}.
\end{equation}
Another effect of increasing $\alpha$ is a dramatic decrease
of the spinon's effective mass
\begin{equation}
m=\left.\left(\frac{\partial^2 \omega_q}{\partial q^2}\right)^{-1}\right|_{q=\pi/2}
\end{equation}
which goes from $(4\alpha J_z)^{-1}$ at $\alpha\ll 1$ to
$\omega_s/c^2$ at $\alpha \agt 0.5$. Such a change can be observed in
the spectra in Fig. \ref{fig:spinon_dispersion}, where increasing $\alpha$ makes
the energy minimum into a sharp tip, indicating the mass
reduction. Thus, even without knowing a specific form of interaction,
one can anticipate that the spinon-holon binding will be strongly
affected by such changes in the spinon spectrum. We also note that
since the spinon becomes much lighter than the holon,
$m_s\ll\!m_h\!\simeq\!(2t)^{-1}$, the role of the holon dispersion
in (\ref{schrod_pair}) becomes secondary close to the isotropic limit.

The remaining question is that of the spinon-holon interaction.  One
can analyze the binding problem in the small-$\alpha$ limit
rigorously. The changes to the holon and the AF GSEs are of order
$O(\alpha^2)$, while the spinon energy changes in the order
$O(\alpha)$: 
\begin{equation}
\omega_q = \omega^0 + \delta\omega_q=
J_z/2+\alpha J_z\cos 2q.
\end{equation}
One of the consequences of non-zero anisotropy is the $\pm\pi/2$
momentum of the GS of the spinon. This immediately implies that the
spinon-holon pairing should result in a bound state with finite
total momentum $P\!=\!\pm\pi/2$, in agreement with the numerical data,
shown in Fig. \ref{fig:pair_dispersion}.  Since the energy of the
system is lowered when the AF domain walls associated with the spinon
and holon pass through each other, the interaction between the two can
be written as a ``contact'' attraction of the strength
$V^0\!=\!-J_z/2$. Using real-space considerations we find that this
leads to a direct relation between interaction in the momentum space
and spinon dispersion.  To the order $O(\alpha)$ the interaction can
be shown to be:
\begin{equation}
\label{int_small_alpha}
V_{q,q'}\!=
\!-\omega^0\!-\!(\delta\omega_q\!+\!\delta\omega_{q'})/2.
\end{equation}
This equation may be used to derive an analytic expression for the
binding energy $\Delta$, exact to the first order in
$\alpha$. Substituting (\ref{int_small_alpha}) into (\ref{schrod_pair})
we get
\begin{equation}
\chi_{\bP}(q)=-\frac{1}{E_q}\left[
A\left(\omega^0+\frac{\delta\omega_q}{2}\right)
+\frac{B}{2}\right],
\end{equation}
where
\begin{eqnarray}
A&=&\sum_{q'}\chi_{\bP}(q'),\\
B&=&\sum_{q'}\delta\omega_{q'}\chi_{\bP}(q'),\\
E_q&\equiv& E-\epsilon_{P-q}-\omega_q.
\end{eqnarray}
After inserting this result for $\chi(q)$ into (\ref{schrod_pair}),
dropping the higher-order terms in $\alpha$, and some algebraic
manipulations, we end up with the following equation for $\Delta$:
\begin{equation}
1=-\frac{J_z}{2}\sum_{q}
\frac{1+2\alpha\cos 2q}{\Delta-(\epsilon_{P-q}-\epsilon_0)-2\alpha(\cos 2q+1)}.
\end{equation}
Further expansion in $\alpha$ and calculation using the ``bare'' holon
energy $\epsilon_k=\epsilon^0_k\equiv 2t\cos k$, yields an expression
for $\Delta$ 
which is valid to order $O(\alpha)$:
\begin{equation}
\label{linear_small}
\Delta=\!\Delta_0(1-C\alpha), 
\end{equation}
with $\Delta_0$ given by Eq. (\ref{ising-delta}) and 
\begin{equation}
C = \frac{2J_z^2}{\sqrt{16t^2+J_z^2}\left(\sqrt{16t^2+J_z^2}-4t\right)}.
\end{equation}
Interestingly, the initial slope of $\Delta(\alpha)/\Delta(0)$ 
depends only weakly on the value of $J_z$: it is bound between 
$C=4$ at $J_z=0$ and $C=2$ at $J_z/t \gg 1$ and varies smoothly between them. 
This linear-$\alpha$ result
(\ref{linear_small}) is shown in Fig. \ref{fig:binding_both} with
dashed lines. It is in extremely close agreement with the numerical
data in the small-$\alpha$ regime.

Having established the form (\ref{int_small_alpha}) of the
spinon-holon interaction for small $\alpha$, we can now try to address
the question of how might the general form of the interaction, valid
for any anisotropy $\alpha$, look like. While there are no strict
analytical arguments for it, we may formulate a number of criteria,
which this form must satisfy. First of all, the interaction $V_{q,q'}$
must be symmetric with respect to momenta,
$V_{q,q'}=V_{q',q}$. Second, it must reproduce the small $\alpha$
limit (\ref{int_small_alpha}) as $\alpha\rightarrow 0$. One can also
anticipate that it should be straightforwardly related to the spinon
energy, similar to Eq. (\ref{int_small_alpha}).

\begin{figure}[b]
\includegraphics*[width=1.0\hsize,scale=1.0]{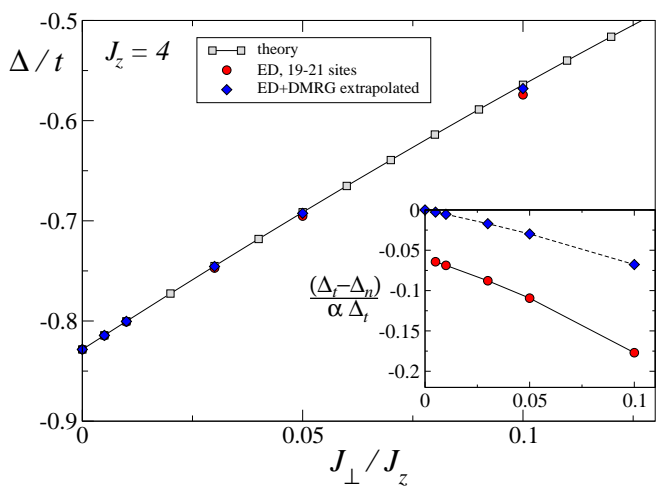}
\caption{(Color online).
Analytic ($\Delta_t$) and numerical ($\Delta_n)$ results for the
binding energy at small values of $\alpha$ for $J_z/t=4.0$. Inset
shows the relative difference between the theoretical and numerical
results.
\label{fig:small_alpha}
}
\end{figure}

Based on these requirements, we propose the following form of the
interaction in the momentum space:
\begin{equation}
\label{vqq_ansatz}
V_{q,q'}=-\sqrt{\omega_q\omega_{q'}}.  
\end{equation}
This is somewhat reminiscent
of the electron-phonon interaction that is proportional to square-root
of the phonon energy. Using this Ansatz for $V_{q,q'}$, spinon energy
from BA, and neglecting the changes in the holon dispersion
($\epsilon_k\!=\!\epsilon^0_k$) we arrive at a solution of
Eq. (\ref{schrod_pair}) of the form
\begin{equation}
\label{chi_a}
\chi(q)=\textrm{const}\times\frac{\sqrt{\omega_q}}{E_q},
\end{equation}
leading to the following equation for $\Delta$:
\begin{equation}
\label{bse_ansatz}
1 =
-\sum_q\frac{\omega_q}{\Delta-
(\epsilon_{P-q}-\epsilon_0)-(\omega_q-\omega_0)}.  
\end{equation}
Solving this equation numerically yields the complete dependence of
the binding energy $\Delta$ on anisotropy $\alpha$ shown as  solid
lines in Fig. \ref{fig:binding_both}.  Not only this
equation naturally yields our small-$\alpha$ results, but it also
provides a very close agreement with the numerical data for all values
of $J_z$ and for all $\alpha$ we can access numerically. This provides
a very convincing \emph{a posteriori} verification of our spinon-holon
interaction Ansatz.

Since we neglect the changes in the holon dispersion, the deviation of
our theoretical result for $\Delta$ from an exact answer is expected
to occur in order $O(\alpha^2)$. We verified that in the
small-$\alpha$ limit. Results of the comparison of analytic and
numerical results are presented in
Fig. \ref{fig:small_alpha}. At $\alpha=0$, 
numerical data obtained using method $B$
for the largest system size accessible by ED ($19$-$21$ sites) agrees
with the exact analytical result (\ref{ising-delta}) 
within the numerical precision ($10^{-7}t$).
However, any small
anisotropy results in a finite-size effect, linear in $\alpha$ (see
inset of Fig. \ref{fig:small_alpha}). This may be understood in terms
of the momentum space mismatch, discussed earlier: for any finite
$\alpha$ and finite size $L$ we cannot obtain the ``true'' GSE value
for spinon from the numerical simulations, because the momentum point
corresponding to its lowest energy ($q=\pi/2$) is
incommensurate with the available momentum points. Thus, a finite-size
effect of the order $O(1/L)$ is expected for any finite $\alpha$. The
extrapolated data, on the other hand, displays the expected
$O(\alpha^2)$ deviation. 

Although we have no formal proof of the validity of our interaction
Ansatz for all $\alpha$, the agreement with the numerical data makes
it very plausible.  As the binding energy becomes small, it is the
long-wavelength features of the dispersions and interaction that
determine the pairing.  One can see from Eq. (\ref{vqq_ansatz}) that
at $\alpha\!\rightarrow\!1$ the characteristic interaction at low
energies is $V\!\approx\!\omega_s$.  Thus, within the qualitative
picture of pairing in 1D, both the interaction and the spinon mass
become proportional to the spinon gap $\omega_s$ that tends to zero
exponentially.  One then expects the asymptotic behavior $\Delta \sim
-V^2m \sim -\omega_s^3$.  From Eq. (\ref{bse_ansatz}) we can derive 
such an asymptotic expression explicitly: $\Delta \approx {\cal
D}(J_z,\alpha)\omega_s^3/c^2$, where
\begin{equation}
{\cal D}(J_z, \alpha) = \left\{ 
\begin{array}{ll}
1, & t\gg J_z;\\
{\displaystyle \frac{1}{2}\left(\frac{\pi c}{4t\ln (c/2t)}\right)^2},
& t\ll J_z. 
\end{array} \right.
\end{equation}
Notably, the exponential behavior of the binding energy is determined
solely by the asymptotic behavior (\ref{gap_asymp}) of the spinon gap
$\omega_s$, with the expression in the exponential dependent only on
$\alpha$ and not on $J_z/t$. Thus, 
the holon energy scale is secondary as it
only enters the prefactor. Altogether, this explains the quick
(exponential) drop-off
\begin{equation}
\Delta\sim -\exp\left(-3\pi^2\sqrt{\frac{\alpha}{8(1-\alpha)}}\right)
\end{equation}
already at intermediate values of $\alpha\gtrsim 0.5$.

From this asymptotic expression and Eq. (\ref{bse_ansatz}) one can see
that the binding energy vanishes in the isotropic limit together with
the spinon gap. Thus, our spinon-holon interaction Ansatz also
provides a natural and simple explanation of the non-zero binding at
finite $q$ but no bound state at $\alpha = 1$. This is possible
because the interaction of the holon with the long-wavelength spinon
$V_{q,q'}$ vanishes together with the spinon energy. Then, the pairing
is not strong enough to produce a bound state in the isotropic limit.
We also find that in the isotropic limit the spinon-holon pair
wave-function, Eq. (\ref{chi_a}), is $\chi(q)\sim 1/\sqrt{\omega_q}$.
In real space, this would correspond to $1/\sqrt{r}$  spinon-holon
correlation, exactly the behavior found in Ref. \onlinecite{Bernevig}.

\section{Conclusions}
\label{sec:conclusions}
We have performed extensive analytical and numerical studies of an
anisotropic version of the $t$-$J$ model, doped with a single
hole. Our main result is that the anisotropy of the spin-spin
interaction leads to an effective attraction between the spinon and
holon excitations, resulting in existence of a spinon-holon bound
state. Using the ED and DMRG techniques we have numerically estimated
the binding energy as a function of anisotropy. We have described in
detail the finite-size effects which arise due to various factors and,
by examining various ways to mitigate or eliminate them, worked out a
procedure for extrapolation of the finite-size data to the infinite
size limit, resulting in precise estimates of the binding energy up to
anisotropy $\alpha=0.5$.  The resulting numerical values have been
found to be in excellent agreement with the theory, based on
Bethe-Salpeter equation. Using the experience gained while studying
the small anisotropy limit, we have formulated the criteria for the
form of the spinon-holon interaction in momentum space, and proposed a
form (\ref{vqq_ansatz}) for it, which results in excellent agreement
of analytical and numerical results. Finally, we have identified the
changes in the spinon spectra as the primary factor affecting the
behavior of the binding energy as a function of anisotropy. We have
demonstrated that the binding energy goes to zero exponentially, as a
power of the spinon gap, when isotropic limit is approached. This
behavior also explains why there is no spinon-holon binding in the
isotropic $t$-$J$ model. These results could be tested in  
photoemission experiments in 1D spin-chain systems with Ising anisotropies.

We would like to note that the problem we have considered is strictly single hole, and its extension to the finite-doping case is not trivial. 
For instance, even in the pure Ising limit one could deliberately avoid creating any spinons by putting an even number of holes in the holon-only states (see Ref. \onlinecite{batista-tjz}). However, if spinons are present in the system, the interaction between them 
and the holons remains attractive even at a finite hole doping and may lead 
to their binding.

\section{Acknowledgments.} 
We would like to thank A. Bernevig and O. Starykh for fruitful
discussions.  This work was supported in part by DOE grant
DE-FG02-04ER46174 and by a Research Corporation Award (J\v{S} and AC) and
by NSF grant DMR-0605444 (SRW).

\appendix
\section{Different treatments of the $\alpha=0$ problem}
For pedagogical purposes we describe three different analytic ways to
determine the bound state energy of the $t$-$J_z$ model, doped with a
single hole.

\emph{Method 1.}\,The first method we discuss is the exact calculation
of the hole's real-space Green's function. It can be accomplished
using the expansion in paths,\cite{nagaoka-hole,brinkman-hole} or
the recursion technique\cite{starykh-hole,sasha} that is also identical
to the Lanczos method. We use the latter as the
most straightforward. The bound state energy may be determined as the
pole of the diagonal element
\begin{equation}
G_{ii}(\omega)=\left\langle \psi_{i}\left|\frac{1}{\omega-{\mathcal
    H}}\right|\psi_{i}\right\rangle, 
\end{equation}
where $\psi_i$ is the state of the system in which the hole is located
at site $i$. 
It may be calculated exactly by noting that we may bring the Hamiltonian
to a tridiagonal form by generating a basis
\begin{equation}
|n+1\rangle=H|n\rangle-a_n|n\rangle-b_n^2|n-1\rangle,
\end{equation}
where
\begin{equation}
a_n=\frac{\langle n|\ham|n\rangle}{\langle n|n\rangle},\quad 
b_n^2=\frac{\langle n|n\rangle}{\langle n-1|n-1 \rangle}.
\end{equation}
It is easy to see that in this basis the diagonal Green's function
may be represented as a continued fraction:
\begin{equation}
\label{g11}
G_{11}(\omega)=\left\langle 1\left|\frac{1}{\omega-\ham}\right|1\right\rangle=
\frac{b_1^2}{\omega-a_1-\frac{b_2^2}{\omega-a_2-\ldots}}.
\end{equation}
In the case of the $t$-$J_z$ model the coefficients in the continued
fraction have the form
\begin{eqnarray}
b_1^2&=&1\nonumber\\
b_2^2&=&2t^2\nonumber\\
b_3^2&=&b_4^2=b_5^2=\ldots=t^2
\end{eqnarray}
and
\begin{eqnarray}
a_1&=& -\wz \equiv -J_z/2\nonumber\\
a_2&=&a_3=a_4=\ldots=0.
\end{eqnarray}
Thus, we may rewrite (\ref{g11}) as
\begin{equation}
G_{ii}(\omega)=\frac{1}{\omega+\wz-2\Sigma(\omega)},
\end{equation}
where
\begin{equation}
\Sigma(\omega)=
\frac{t^2}{\omega-\frac{t^2}{\omega-\ldots}}
=\frac{t^2}{\omega-\Sigma_(\omega)}. 
\end{equation}
Solving for $\Sigma(\omega)$ yields the Green's function
\begin{equation}
\label{gg}
G(\omega)=\frac{1}{J_z/2\mp \sqrt{\omega^2-4t^2}},
\end{equation}
with poles given by
\begin{equation}
\label{gpoles}
\bar{\omega}=\pm\sqrt{4t^2+J_z^2/4}.
\end{equation}
The energy corresponding to the lowest pole is lower
than hole's kinetic energy $-2t$, indicating the presence of a bound
state. The expression for the binding energy
$\Delta=2t - |\bar{\omega}|$ is equivalent to (\ref{ising-delta}).
The quasiparticle residue of that pole is given by:\cite{sorella-1d}
\begin{equation}
\label{qpresidue}
Z=\frac{J_z}{2|\bar{\omega}|}=\frac{J_z}{\sqrt{16t^2+J_z^2}}.
\end{equation}
At large $t/J_z$, in agreement with the naive expectations, $Z\approx (J_z/2)/2t$, the ratio of spinon-holon interaction strength to the holon kinetic energy. 

\emph{Method 2.}\,Another approach is to consider the immobile spinon
to be an impurity 
and solve the problem of a freely moving hole scattering on it using
the $T$-matrix 
formalism. To that end we write the Hamiltonian of the hole as
\begin{equation}
\ham_0=-2t\sum_k\cos(k)c^{\dagger}_kc_k,
\end{equation} 
where $c_k$ is the annihilation operator for a hole with momentum $k$.
The impurity Hamiltonian may be written as
\begin{equation}
\ham'=-\wz c^{\dagger}_rc_r,
\end{equation}
i.e. it lowers the energy by $\omega^0\equiv J_z/2$ if the hole is
present at the origin 
site. Transforming it to the momentum space after assuming $r=0$ yields
\begin{equation}
\ham'=-\wz\sum_{k,k'}c^{\dagger}_kc_{k'}.
\end{equation} 
The impurity Hamiltonian is such that each of its matrix elements is
equal to $-\wz$. The equation for $T$-matrix is
\begin{equation}
T=\ham'+\ham'G_0T.
\end{equation}
Inserting complete sets of states and using the fact that $G_0$ is diagonal and
$\langle k|\ham'|k'\rangle=-\wz$ for arbitrary $k$, $k'$, we get the
following equation for the matrix elements:
\begin{equation}
\langle k|T|k'\rangle=-\wz-\wz\sum_p\langle p|G_0|p\rangle\langle
p|T|k'\rangle. 
\end{equation}
Close examination of this expression reveals that the $T$-matrix is
independent of $k$ and $k'$. Denoting the matrix element by
$T(\omega)$ we obtain
\begin{equation}
T(\omega)=-\wz \left(1+T(\omega)\sum_p \langle p|G_0(\omega)|p\rangle\right).
\end{equation}
The integral on the rhs
\begin{equation}
S=\sum_p \langle p|G_0(\omega)|p\rangle
\end{equation}
can be calculated using elementary methods to find
\begin{equation}
S=\frac{1}{\sqrt{\omega^2-4t^2}}.
\end{equation}
Thus
\begin{equation}
T(\omega)=-\frac{\wz\sqrt{\omega^2-4t^2}}{\wz+\sqrt{\omega^2-4t^2}}
\end{equation}
The exact Green's function of a particle with a scatterer is a function
of incoming and outgoing momenta $k$ and $k'$, given by
\begin{equation}
G(k, k')=G_0(k)+G_0(k)TG_0(k').
\end{equation}
This yields the following matrix element:
\begin{equation}
\langle k|G(\omega)|k'\rangle=\frac{1}{\omega+2t\cos k}\left[\delta_{k,k'}
+T(\omega)\frac{1}{\omega+2t\cos k'}\right].
\end{equation}
Transforming to real space by integrating over momenta $k$ and $k'$,
we get the diagonal Green's function at the point of origin:
\begin{eqnarray}
\label{gg_r}
G_{rr}(\omega)&=&\sum_{k,k'}\langle k|G(\omega)|k'\rangle\nonumber\\
&=&\frac{1}{\sqrt{\omega^2-4t^2}}+\frac{1}{\omega^2-4t^2}T(\omega)\nonumber\\
&=&\frac{1}{\wz+\sqrt{\omega^2-4t^2}}
\end{eqnarray}
that is equivalent to (\ref{gg}).
This Green's function has the poles at the locations given by (\ref{gpoles}), 
so it yields binding energy  equivalent to (\ref{ising-delta}).

\emph{Method 3.}\,Finally, the last approach is based on the physical
picture of hole decay into a spinon and holon, confined to two
half-spaces (see Fig. \ref{fig:spinon_holon}), and solving the Dyson's
equation for such a decay exactly. The Dyson equation for the hole at
origin has the form
\begin{equation}
\label{gfholon}
G(\omega)=\frac{1}{(G^0)^{-1}-\Sigma(\omega)},
\end{equation} 
where
\begin{equation}
G^0(\omega)=\frac{1}{\omega},
\end{equation}
and self-energy $\Sigma(\omega)$ may
be written in terms of the spinon Green's function $D(\omega)$ and the
holon Green's function in a half-space $G^{(1/2)}_h(\omega)$:
\begin{equation}
\Sigma(\omega)=2t^2\int\,
\frac{d\omega'}{2\pi}D(\omega')G^{(1/2)}_h(\omega-\omega'). 
\end{equation}
Since 
\begin{equation}
D(\omega)=\frac{1}{\omega-\wz},
\end{equation}
the integral is readily done, leading to the expression for self-energy
\begin{equation}
\Sigma(\omega)=2t^2G^{(1/2)}_h(\omega-\wz).
\end{equation}
The Green's function (\ref{gfholon}) can then be written in terms of
the shifted frequency 
$\tilde{\omega}=\omega-\wz$ as
\begin{equation}
G(\tilde{\omega})=\frac{1}{\tilde{\omega}+\wz-\Sigma(\tilde{\omega})}.
\end{equation}
To calculate $G_h^{(1/2)}(\tilde\omega)$, that is the Green's function of a
free hole subject to a ``hard-wall'' boundary condition at the origin,
we note that it can be expressed as the antisymmetric part of the
hole's Green's function $G_h(\omega)$ in the entire space. For
example, in the coordinate representation we obtain:
\begin{equation}
\label{g-half}
G_h^{(1/2)}(x, x'; \omega)=G_h(x, x'; \omega)-G_h(x, -x'; \omega).
\end{equation}
Here $G_h(x, x'; \omega)$ is just an inverse Fourier transform of
the Green's function of a free hole:
\begin{equation}
\label{g-whole}
G_h(x, x';\omega)=\int\frac{dp}{2\pi}\frac{e^{i(x-x')p}}{\omega+2t\cos p}.
\end{equation}
It depends only on the difference of the coordinates $x-x'$, as expected
for a translationally invariant system. We are interested in the diagonal
matrix element of (\ref{g-half}):
\begin{eqnarray}
G_h^{(1/2)}(x; \omega)&\equiv& G_h^{(1/2)}(x, x; \omega)=\nonumber\\
&=&G_h(x-x; \omega)-G_h(x+x; \omega).
\end{eqnarray}
Using (\ref{g-whole}) it may be readily found to be
\begin{equation}
2G_h^{(1/2)}(\omega)=\omega\pm\sqrt{\omega^2-4t^2},
\end{equation}
again leading to the expression for $G(\omega)$ equivalent to
 Eqs. (\ref{gg}) and (\ref{gg_r}) and expression for 
 $\Delta$, equivalent to (\ref{ising-delta}). 
\bibliography{refs}

\begin{thebibliography}{25}
\expandafter\ifx\csname natexlab\endcsname\relax\def\natexlab#1{#1}\fi
\expandafter\ifx\csname bibnamefont\endcsname\relax
  \def\bibnamefont#1{#1}\fi
\expandafter\ifx\csname bibfnamefont\endcsname\relax
  \def\bibfnamefont#1{#1}\fi
\expandafter\ifx\csname citenamefont\endcsname\relax
  \def\citenamefont#1{#1}\fi
\expandafter\ifx\csname url\endcsname\relax
  \def\url#1{\texttt{#1}}\fi
\expandafter\ifx\csname urlprefix\endcsname\relax\def\urlprefix{URL }\fi
\providecommand{\bibinfo}[2]{#2}
\providecommand{\eprint}[2][]{\url{#2}}

\bibitem[{\citenamefont{Giamarchi}(2004)}]{giamarchi-1d}
\bibinfo{author}{\bibfnamefont{T.}~\bibnamefont{Giamarchi}},
  \emph{\bibinfo{title}{Quantum Physics in One Dimension}}, vol.
  \bibinfo{volume}{121} of \emph{\bibinfo{series}{International Series of
  Monographs on Physics}} (\bibinfo{publisher}{Oxford University Press},
  \bibinfo{year}{2004}).

\bibitem[{\citenamefont{Lieb and Wu}(1968)}]{LiebWu}
\bibinfo{author}{\bibfnamefont{E.~H.} \bibnamefont{Lieb}} \bibnamefont{and}
  \bibinfo{author}{\bibfnamefont{F.~Y.} \bibnamefont{Wu}},
  \bibinfo{journal}{Phys. Rev. Lett.} \textbf{\bibinfo{volume}{20}},
  \bibinfo{pages}{1445} (\bibinfo{year}{1968}).

\bibitem[{\citenamefont{Shiba and Ogata}(1992)}]{shiba-ogata-review}
\bibinfo{author}{\bibfnamefont{H.}~\bibnamefont{Shiba}} \bibnamefont{and}
  \bibinfo{author}{\bibfnamefont{M.}~\bibnamefont{Ogata}},
  \bibinfo{journal}{Prog. Theor. Phys. Supp.} \textbf{\bibinfo{volume}{108}},
  \bibinfo{pages}{265} (\bibinfo{year}{1992}).

\bibitem[{\citenamefont{Sorella and Parola}(1998)}]{sorella-1d}
\bibinfo{author}{\bibfnamefont{S.}~\bibnamefont{Sorella}} \bibnamefont{and}
  \bibinfo{author}{\bibfnamefont{A.}~\bibnamefont{Parola}},
  \bibinfo{journal}{Phys. Rev. B} \textbf{\bibinfo{volume}{57}},
  \bibinfo{pages}{6444} (\bibinfo{year}{1998}).

\bibitem[{\citenamefont{Brunner et~al.}(2000)\citenamefont{Brunner, Assaad, and
  Muramatsu}}]{muramatsu}
\bibinfo{author}{\bibfnamefont{M.}~\bibnamefont{Brunner}},
  \bibinfo{author}{\bibfnamefont{F.~F.} \bibnamefont{Assaad}},
  \bibnamefont{and}
  \bibinfo{author}{\bibfnamefont{A.}~\bibnamefont{Muramatsu}},
  \bibinfo{journal}{Eur. Phys. J. B} \textbf{\bibinfo{volume}{16}},
  \bibinfo{pages}{209} (\bibinfo{year}{2000}).

\bibitem[{\citenamefont{Bernevig et~al.}(2002)\citenamefont{Bernevig, Giuliano,
  and Laughlin}}]{Bernevig}
\bibinfo{author}{\bibfnamefont{B.~A.} \bibnamefont{Bernevig}},
  \bibinfo{author}{\bibfnamefont{D.}~\bibnamefont{Giuliano}}, \bibnamefont{and}
  \bibinfo{author}{\bibfnamefont{R.~B.} \bibnamefont{Laughlin}},
  \bibinfo{journal}{Phys. Rev. B} \textbf{\bibinfo{volume}{65}},
  \bibinfo{pages}{195112} (\bibinfo{year}{2002}).

\bibitem[{\citenamefont{Bares et~al.}(1991)\citenamefont{Bares, Blatter, and
  Ogata}}]{bares-tj-exact-2t}
\bibinfo{author}{\bibfnamefont{P.-A.} \bibnamefont{Bares}},
  \bibinfo{author}{\bibfnamefont{G.}~\bibnamefont{Blatter}}, \bibnamefont{and}
  \bibinfo{author}{\bibfnamefont{M.}~\bibnamefont{Ogata}},
  \bibinfo{journal}{Phys. Rev. B} \textbf{\bibinfo{volume}{44}},
  \bibinfo{pages}{130} (\bibinfo{year}{1991}).

\bibitem[{\citenamefont{Nagler et~al.}(1983)\citenamefont{Nagler, Buyers,
  Armstrong, and Briat}}]{nagler-1d-ising}
\bibinfo{author}{\bibfnamefont{S.~E.} \bibnamefont{Nagler}},
  \bibinfo{author}{\bibfnamefont{W.~J.~L.} \bibnamefont{Buyers}},
  \bibinfo{author}{\bibfnamefont{R.~L.} \bibnamefont{Armstrong}},
  \bibnamefont{and} \bibinfo{author}{\bibfnamefont{B.}~\bibnamefont{Briat}},
  \bibinfo{journal}{Phys. Rev. B} \textbf{\bibinfo{volume}{27}},
  \bibinfo{pages}{1784} (\bibinfo{year}{1983}).

\bibitem[{\citenamefont{Kim et~al.}(1996)\citenamefont{Kim, Matsuura, Shen,
  Motoyama, Eisaki, Uchida, Tohyama, and Maekawa}}]{kim-separation-srcuo2}
\bibinfo{author}{\bibfnamefont{C.}~\bibnamefont{Kim}},
  \bibinfo{author}{\bibfnamefont{A.~Y.} \bibnamefont{Matsuura}},
  \bibinfo{author}{\bibfnamefont{Z.~X.} \bibnamefont{Shen}},
  \bibinfo{author}{\bibfnamefont{N.}~\bibnamefont{Motoyama}},
  \bibinfo{author}{\bibfnamefont{H.}~\bibnamefont{Eisaki}},
  \bibinfo{author}{\bibfnamefont{S.}~\bibnamefont{Uchida}},
  \bibinfo{author}{\bibfnamefont{T.}~\bibnamefont{Tohyama}}, \bibnamefont{and}
  \bibinfo{author}{\bibfnamefont{S.}~\bibnamefont{Maekawa}},
  \bibinfo{journal}{Phys. Rev. Lett.} \textbf{\bibinfo{volume}{77}},
  \bibinfo{pages}{4054} (\bibinfo{year}{1996}).

\bibitem[{\citenamefont{\v{S}makov et~al.}(2007)\citenamefont{\v{S}makov,
  Chernyshev, and White}}]{short}
\bibinfo{author}{\bibfnamefont{J.}~\bibnamefont{\v{S}makov}},
  \bibinfo{author}{\bibfnamefont{A.~L.} \bibnamefont{Chernyshev}},
  \bibnamefont{and} \bibinfo{author}{\bibfnamefont{S.~R.} \bibnamefont{White}},
  \bibinfo{journal}{Phys. Rev. Lett.} \textbf{\bibinfo{volume}{98}},
  \bibinfo{pages}{266401} (\bibinfo{year}{2007}).

\bibitem[{\citenamefont{Johnson et~al.}(1973)\citenamefont{Johnson, Krinsky,
  and McCoy}}]{johnson-excitations-xyz}
\bibinfo{author}{\bibfnamefont{J.~D.} \bibnamefont{Johnson}},
  \bibinfo{author}{\bibfnamefont{S.}~\bibnamefont{Krinsky}}, \bibnamefont{and}
  \bibinfo{author}{\bibfnamefont{B.~M.} \bibnamefont{McCoy}},
  \bibinfo{journal}{Phys. Rev. A} \textbf{\bibinfo{volume}{8}},
  \bibinfo{pages}{2526} (\bibinfo{year}{1973}).

\bibitem[{\citenamefont{Haas}(1995)}]{haas}
\bibinfo{author}{\bibfnamefont{S.}~\bibnamefont{Haas}}, Ph.D. thesis,
  \bibinfo{school}{Florida State University} (\bibinfo{year}{1995}),
  \urlprefix\url{http://physics.usc.edu/~shaas/haasthesis.pdf}.

\bibitem[{\citenamefont{White}(1992)}]{white-one}
\bibinfo{author}{\bibfnamefont{S.~R.} \bibnamefont{White}},
  \bibinfo{journal}{Phys. Rev. Lett.} \textbf{\bibinfo{volume}{69}},
  \bibinfo{pages}{2863} (\bibinfo{year}{1992}).

\bibitem[{\citenamefont{White}(1993)}]{white-two}
\bibinfo{author}{\bibfnamefont{S.~R.} \bibnamefont{White}},
  \bibinfo{journal}{Phys. Rev. B.} \textbf{\bibinfo{volume}{48}},
  \bibinfo{pages}{10345} (\bibinfo{year}{1993}).

\bibitem[{\citenamefont{White}(2005)}]{white-three}
\bibinfo{author}{\bibfnamefont{S.~R.} \bibnamefont{White}},
  \bibinfo{journal}{Phys. Rev. B.} \textbf{\bibinfo{volume}{72}},
  \bibinfo{pages}{180403} (\bibinfo{year}{2005}).

\bibitem[{\citenamefont{Medeiros and Cabrera}(1991)}]{medeiros-size-corr}
\bibinfo{author}{\bibfnamefont{D.}~\bibnamefont{Medeiros}} \bibnamefont{and}
  \bibinfo{author}{\bibfnamefont{G.~G.} \bibnamefont{Cabrera}},
  \bibinfo{journal}{Phys. Rev. B} \textbf{\bibinfo{volume}{44}},
  \bibinfo{pages}{848} (\bibinfo{year}{1991}).

\bibitem[{\citenamefont{Yang and Yang}(1966)}]{yang-xxz}
\bibinfo{author}{\bibfnamefont{C.~N.} \bibnamefont{Yang}} \bibnamefont{and}
  \bibinfo{author}{\bibfnamefont{C.~P.} \bibnamefont{Yang}},
  \bibinfo{journal}{Phys. Rev.} \textbf{\bibinfo{volume}{150}},
  \bibinfo{pages}{321} (\bibinfo{year}{1966}).

\bibitem[{\citenamefont{de~Vega and Woynarovich}(1985)}]{vega-woynar-size-corr}
\bibinfo{author}{\bibfnamefont{H.~J.} \bibnamefont{de~Vega}} \bibnamefont{and}
  \bibinfo{author}{\bibfnamefont{F.}~\bibnamefont{Woynarovich}},
  \bibinfo{journal}{Nucl. Phys. B} \textbf{\bibinfo{volume}{251}},
  \bibinfo{pages}{439} (\bibinfo{year}{1985}).

\bibitem[{\citenamefont{Zotos et~al.}(1990)\citenamefont{Zotos, Prelovek, and
  Sega}}]{zotos-tbc}
\bibinfo{author}{\bibfnamefont{X.}~\bibnamefont{Zotos}},
  \bibinfo{author}{\bibfnamefont{P.}~\bibnamefont{Prelovek}}, \bibnamefont{and}
  \bibinfo{author}{\bibfnamefont{I.}~\bibnamefont{Sega}},
  \bibinfo{journal}{Phys. Rev. B} \textbf{\bibinfo{volume}{42}},
  \bibinfo{pages}{8445} (\bibinfo{year}{1990}).

\bibitem[{\citenamefont{Berestetskii et~al.}(1982)\citenamefont{Berestetskii,
  Lifshitz, and Pitaevskii}}]{landau-qed}
\bibinfo{author}{\bibfnamefont{V.~B.} \bibnamefont{Berestetskii}},
  \bibinfo{author}{\bibfnamefont{E.~M.} \bibnamefont{Lifshitz}},
  \bibnamefont{and} \bibinfo{author}{\bibfnamefont{L.~P.}
  \bibnamefont{Pitaevskii}}, \emph{\bibinfo{title}{Quantum Electrodynamics}},
  vol.~\bibinfo{volume}{4} of \emph{\bibinfo{series}{Landau and Lifshitz Course
  of Theoretical Physics}} (\bibinfo{publisher}{Pergamon Press},
  \bibinfo{year}{1982}).

\bibitem[{\citenamefont{Batista and Ortiz}(2000)}]{batista-tjz}
\bibinfo{author}{\bibfnamefont{C.~D.} \bibnamefont{Batista}} \bibnamefont{and}
  \bibinfo{author}{\bibfnamefont{G.}~\bibnamefont{Ortiz}},
  \bibinfo{journal}{Phys. Rev. Lett.} \textbf{\bibinfo{volume}{85}},
  \bibinfo{pages}{4755} (\bibinfo{year}{2000}).

\bibitem[{\citenamefont{Nagaoka}(1965)}]{nagaoka-hole}
\bibinfo{author}{\bibfnamefont{Y.}~\bibnamefont{Nagaoka}},
  \bibinfo{journal}{Sol. State Comm.} \textbf{\bibinfo{volume}{3}},
  \bibinfo{pages}{409} (\bibinfo{year}{1965}).

\bibitem[{\citenamefont{Brinkman and Rice}(1970)}]{brinkman-hole}
\bibinfo{author}{\bibfnamefont{W.~F.} \bibnamefont{Brinkman}} \bibnamefont{and}
  \bibinfo{author}{\bibfnamefont{T.~M.} \bibnamefont{Rice}},
  \bibinfo{journal}{Phys. Rev. B} \textbf{\bibinfo{volume}{2}},
  \bibinfo{pages}{1324} (\bibinfo{year}{1970}).

\bibitem[{\citenamefont{Starykh and Reiter}(1996)}]{starykh-hole}
\bibinfo{author}{\bibfnamefont{O.~A.} \bibnamefont{Starykh}} \bibnamefont{and}
  \bibinfo{author}{\bibfnamefont{G.~F.} \bibnamefont{Reiter}},
  \bibinfo{journal}{Phys. Rev. B} \textbf{\bibinfo{volume}{53}},
  \bibinfo{pages}{2517} (\bibinfo{year}{1996}).

\bibitem[{\citenamefont{Chernyshev and Leung}(1999)}]{sasha}
\bibinfo{author}{\bibfnamefont{A.~L.} \bibnamefont{Chernyshev}}
  \bibnamefont{and} \bibinfo{author}{\bibfnamefont{P.~W.} \bibnamefont{Leung}},
  \bibinfo{journal}{Phys. Rev. B.} \textbf{\bibinfo{volume}{60}},
  \bibinfo{pages}{1592} (\bibinfo{year}{1999}).

\end{thebibliography}
\pagebreak

\end{document}